\documentclass[12pt]{iopart}

\usepackage[utf8]{inputenc}
\usepackage{epsfig}
\usepackage[T1]{fontenc}
\usepackage{graphicx}
\usepackage{amssymb}
\usepackage{amsmath}
\usepackage{cite}

\usepackage{relsize}
\usepackage{color}
\usepackage{bm}
\usepackage{array}
\newcommand{\avg}[1]{{\left<#1\right>}}
\newcommand{\floor}[1]{{\lfloor #1\rfloor}}
\newcommand{\ceil}[1]{{\lceil #1\rceil}}

\begin{document}

\title{The behavior of noise-resilient Boolean networks with diverse
  topologies}

\author{Tiago P. Peixoto}
\address{Institut für Theoretische Physik, Universität Bremen,
Otto-Hahn-Allee 1, D-28359 Bremen, Germany}
\ead{tiago@itp.uni-bremen.de}

\begin{abstract}
  The dynamics of noise-resilient Boolean networks with majority
  functions and diverse topologies is investigated. A wide class of
  possible topological configurations is parametrized as a stochastic
  blockmodel. For this class of networks, the dynamics always undergoes
  a phase transition from a non-ergodic regime, where a memory of its
  past states is preserved, to an ergodic regime, where no such memory
  exists and every microstate is equally probable. Both the average
  error on the network, as well as the critical value of noise where the
  transition occurs are investigated analytically, and compared to
  numerical simulations. The results for ``partially dense'' networks,
  comprised of relatively few, but dynamically important nodes, which
  have a number of inputs which greatly exceeds the average for the
  entire network, give very general upper bounds on the maximum
  resilience against noise attainable on globally sparse systems.
\end{abstract}


\maketitle

\section{Introduction}

An essential feature of many self-organized and artificial systems of
several interacting elements is the ability of to function in a
predictable fashion even in the presence of stochastic fluctuations,
which are inherent to the system itself. Good examples are biochemical
signaling networks and gene regulation in
organisms~\cite{kitano_biological_2004}, as well as artificial digital
circuits, and communication
networks~\cite{shannon_mathematical_1948}. In such systems, it is often
the case that the source of the fluctuations cannot be entirely removed,
and the system must be able to deal with them, by incorporating
appropriate error-correction measures. These may include specific
dynamical properties~\cite{peixoto_boolean_2009,schmal_boolean_2010},
choice of functional elements and structural
properties~\cite{peixoto_redundancy_2010, peixoto_emergence_2011}, which
one way or another result in enough information redundancy, which can be
used to counteract the deviating effects of noise. In this work, the
focus is turned on optimal bounds which can be attained by a wide class
of such systems, when many parameters can be freely varied. More
precisely, we consider a paradigmatic system of dynamically interacting
Boolean elements, regulated by Boolean functions, where noise is
introduced by the probability that at any time, any input of a given
function can be ``flipped'' to its opposite value, before the output of
the function is computed. The networks considered are regulated by
optimal majority functions, and can possess arbitrary topological
structures. The choice of majority functions corresponds to the limiting
case where the trade-off between robustness against noise and fitness
for a given task is at a maximum for every function in the network.

We obtain -- both analytically and numerically -- relevant properties of
the system, such as the average probability of error as a function of
noise, and critical value of noise, for which reliability is no longer
possible. At this noise threshold, the system undergoes a dynamic phase
transition from a non-ergodic regime, where a memory of its past states
is preserved, to an ergodic regime, where no such memory exists and
every microstate is equally probable. We identify the most relevant
topological properties which can confer more robustness to the system,
namely the existence of a more densely connected subset of the network,
which is responsible for the dynamics of a significant portion of the
system. The properties of such optimal topologies serve as general
optimal bounds on the maximum resilience against noise which is
attainable by this class of system.

The behavior of similar systems under noise has been studied previously
by a number of authors. The dynamics of random Boolean networks (RBNs)
with noise (random functions and topology, not necessarily aiming at
robustness~\cite{drossel_random_2008-1}) was studied in~\cite{
  miranda_noise_1989, golinelli_barrier_1989, qu_numerical_2002,
  huepe_dynamical_2002, aleksiejuk_ferromagnetic_2002,
  indekeu_special_2004, fretter_perturbation_2009,
  peixoto_noise_2009}. The early works presented
in~\cite{miranda_noise_1989,golinelli_barrier_1989, qu_numerical_2002}
considered only small networks with $N \le 20$ nodes, and focused on the
average crossing time between trajectories in state space which started
from different initial states. It was found that the trajectories must
cross over ``barriers,'' which correspond to the attractor basin
boundaries. However, the probability of crossing is always non-vanishing
in such small systems. It was further shown in~\cite{peixoto_noise_2009}
that the dynamics of RBNs is always ergodic for any positive value of
noise, and thus cannot preserve any memory of its past states. However,
this is not true for random networks composed of threshold or majority
functions, as shown
in~\cite{huepe_dynamical_2002,aldana_phase_2004}. These networks undergo
the aforementioned phase transition between ergodicity and
non-ergodicity at a critical value of noise. The same type of transition
has also been observed for Boolean systems composed of majority
functions, but having acyclic and stratified topology (i.e. Boolean
\emph{formulas})~\cite{mozeika_computing_2009, mozeika_noisy_2010}. It
was also shown in~\cite{peixoto_redundancy_2010} that this transition
has a general character, since any Boolean network can be made robust by
introducing an appropriate restoration mechanism with majority
functions.

Boolean networks with majority functions share some similarities with
the so-called majority voter
model~\cite{oliveira_isotropic_1992,oliveira_nonequilibrium_1993}, which
is usually defined on undirected regular lattices. This system also
undergoes a phase-transition based on noise, which belongs to the
universality class of the Ising model~\cite{grinstein_statistical_1985}.

The issue of reliable computation under noise has also been tackled by
the mathematical community, starting with von
Neumann~\cite{von_neumann_probabilistic_1956}, who was the first to
notice an important difference between reliable computation of noisy
Boolean circuits and the more general scenario of reliable communication
considered by Shannon~\cite{shannon_mathematical_1948}, namely that it
is not possible to guarantee an arbitrarily small error rate, if the a
given circuit has a fixed number of inputs per function. He also pointed
out that reliable computation is not at all possible for Boolean
functions with three inputs after a given noise threshold. His results
were later improved by Evans and Pippinger~\cite{evans_maximum_1998},
who proved a similar bound for Boolean formulas with two inputs per
node, and finally Evans and Schulman~\cite{evans_maximum_2003} who
proved the bound for Boolean formulas with any odd number of inputs per
node.  Recently, an extension to these bounds which are also valid for
functions with even number of inputs was derived
in~\cite{mozeika_dynamics_2011}.

This paper is divided as follows. In section~\ref{sec:model} we describe
the model and in section~\ref{sec:transition} we analyse the phase
transition based on noise for several different topological models:
In~\ref{sec:random} we consider random networks with a single-valued
in-degree distribution, and in~\ref{sec:deg-dist} we extend the model to
arbitrary in-degree distributions. In~\ref{sec:block} we consider a more
general stochastic blockmodel, which represents a much larger class of
possible topological structures. We finalize in
section~\ref{sec:conclusion} with some concluding remarks.

\section{The model}\label{sec:model}

A Boolean Network (BN)~\cite{kauffman_metabolic_1969,drossel_random_2008}
is a directed graph of $N$ nodes representing Boolean variables
$\mathbf{\sigma} \in \{1,0\}^N$, which are subject to a deterministic
update rule,
\begin{equation}\label{eq:bn_dyn}
  \sigma_i(t+1) = f_i\left(\bm{\sigma}(t)\right)
\end{equation}
where $f_i$ is the update function assigned to node $i$, which depends
exclusively on the states of its inputs. It is also considered that all
nodes are updated in parallel.

Here, noise is included in the model by introducing a probability $P$
that at each time-step a given input has its value ``flipped'':
$\sigma_j \to 1 - \sigma_j$, before the output is
computed~\cite{peixoto_noise_2009}. This probability is independent for
all inputs in the network, and many values may be flipped
simultaneously. The functions on all nodes are taken to be the majority
function, defined as
\begin{equation}\label{eq:maj_f}
  f_i(\{\sigma_j\}) = 
  \begin{cases}
    1  \text{ if } \sum_j\sigma_j > k_i / 2, \\
    0  \text{ otherwise, }
  \end{cases}
\end{equation}
where $k_i$ is the number of inputs of node $i$. The definition above
will lead to a bias if $k_i$ is an even number, since if the sum happens
to be exactly $k_i / 2$ the output will be $0$, arbitrarily. Alternative
definitions could be used, which would remove the
bias~\cite{szejka_phase_2008}. Instead, for the sake of simplicity, in
this work all values of $k_i$ considered will be odd, making this bias a
non-issue.

Starting from a given initial configuration, the dynamics of the system
evolves and eventually reaches a dynamically stable regime, where (for
sufficiently large systems) the average value $b_t$ of $1$'s no longer
changes, except for stochastic
fluctuations~\cite{huepe_dynamical_2002}. In the absence of noise
($P=0$) there are only two possible attractors (if the network is
sufficiently random and not disjoint), where all nodes have the same
value, which can be either $0$ or $1$. We will consider these
homogeneous attractors as being the ``correct'' dynamics, and denote the
deviations from them as ``errors''. More specifically, without loss of
generality, we will name the value of $1$ as an ``error'', and the value
of $b_t$ as the average error on the system.

We note that the above model has an optimal character regarding
robustness against noise, for the following two reasons: 1. It is known
that the majority function as defined in Eq.~\ref{eq:maj_f} is optimal
in the case of fully redundant inputs (i.e. in the absence of noise,
they all have simultaneously the same value), which have an uniform and
independent probability of being ``flipped'' by noise. In this
situation, the output of the majority function will be ``correct'' with
greater probability than any other function with the same number of
inputs~\cite{von_neumann_probabilistic_1956, evans_maximum_2003}. 2. The
existence of only two possible attractors with uniform values can be
interpreted as an extremal trade-off between dynamical function and
increased resilience against noise: A network with more elaborate
dynamics in the absence of noise, composed of many attractors with
smaller basis of attraction, would be invariably more difficult to
stabilize if noise is present, since it would become harder to
distinguish between dynamical states.

\section{Dynamical phase transition based on noise}
\label{sec:transition}

As previously defined, the average ``error'' on the network is
characterized by the average value of $1$'s in the network at a given
time, $b_t$. In this section we will obtain the value of $b_t$ for
networks with different topological characteristics. We will focus first
on uniform random networks with all functions having the same in-degree,
and networks with arbitrary in-degree distributions. We then move to an
arbitrary blockmodel, which can incorporate more general topological
features.

\subsection{Single-valued in-degree distribution}\label{sec:random}

In this session, we compute the value of $b_t$ for networks composed of
nodes with the same number $k$ of inputs per node, which are randomly
chosen between all possible nodes. This type of system has been studied
before by Huepe et al~\cite{huepe_dynamical_2002} and is essentially
equivalent to the same problem posed for Boolean \emph{formulas} by
Evans et. al~\cite{evans_maximum_2003}, since the presence of short
loops can be neglected for large networks. For the sake of clarity, we
shortly reproduce the analysis developed in~\cite{evans_maximum_2003},
and we extend it by calculating the critical exponent of the
transition. We then proceed to generalize the approach to more general
topologies in the subsequent sections.

\begin{figure}[hbt!]
  \includegraphics*[width=0.49\columnwidth]{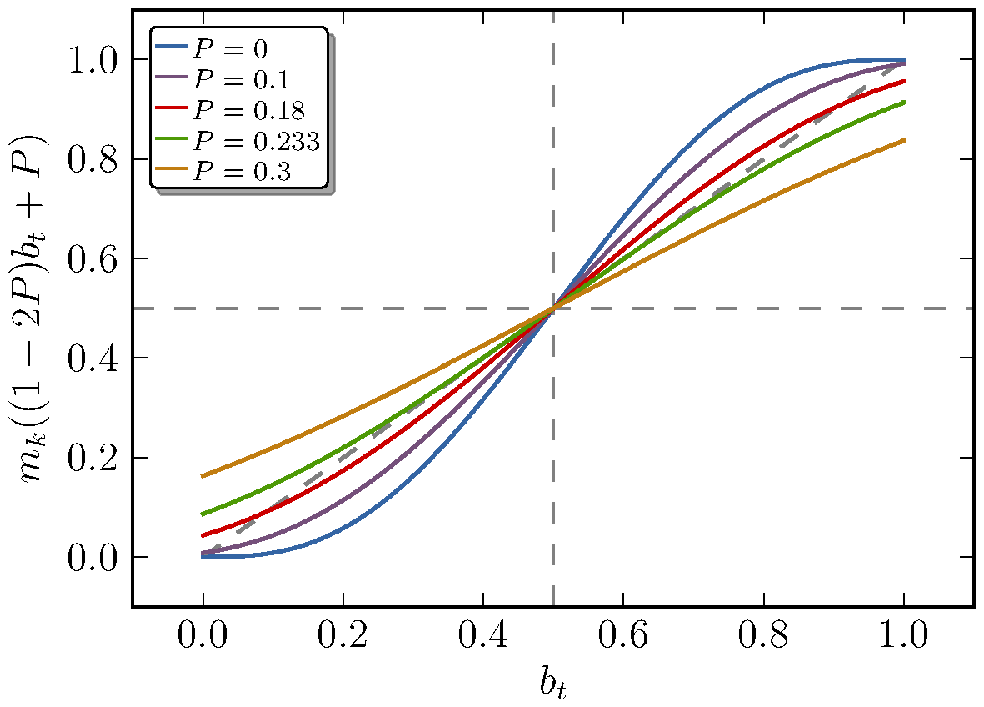}
  \includegraphics*[width=0.49\columnwidth]{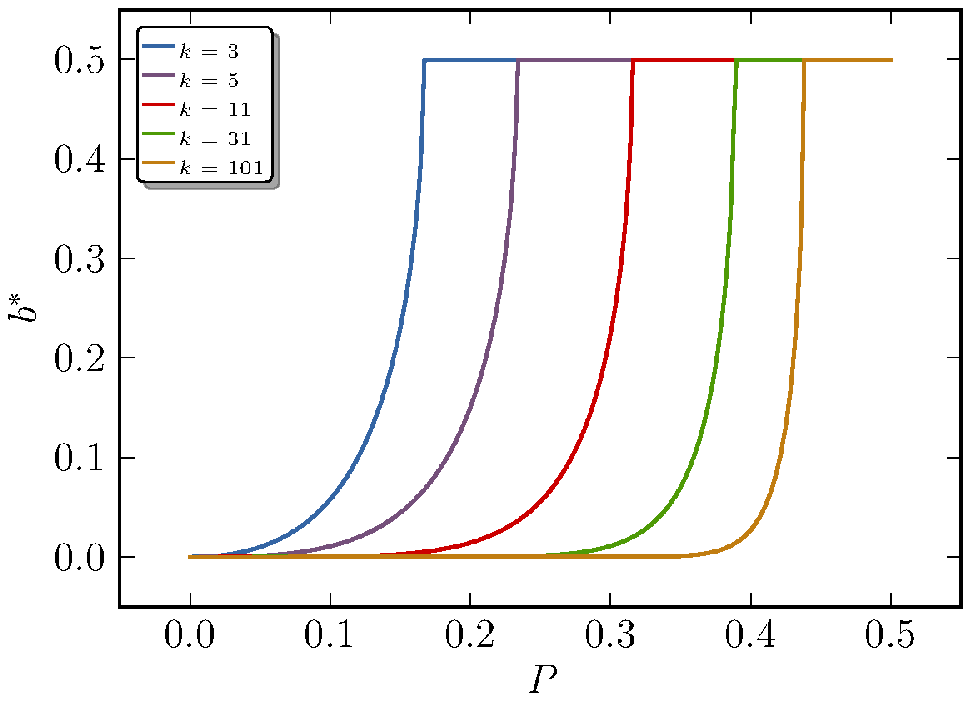}
  \caption{The dynamic map of Eq.~\ref{eq:dyn_fk} for different values
    of $P$ (left), and the value of the stable fixed-point $b^* \le
    1/2$, as a function of $P$ (right).\label{fig:maj-dyn}}
\end{figure}

In order to obtain an equation for the time evolution of $b_t$ we employ
the usual annealed approximation~\cite{derrida_random_1986}, which
assumes that at each time step the inputs of every function are randomly
re-sampled, such that any quenched topological correlations are ignored,
and all inputs will have the same probability $b_t$ of being equal to
$1$. If the inputs of a majority function have a value of $1$ with
probability $b$ (independently for each input), the output will also be
$1$ with a probability given by
\begin{equation}\label{eq:maj}
  m_k(b) = \sum_{i = \ceil{k/2}}^k{k \choose i} b^i(1-b)^{k-i}.
\end{equation}
The time evolution of $b_t$ can then be written as
\begin{equation}\label{eq:dyn_fk}
  b_{t+1} = m_k\left((1-2P) b_{t} + P\right),
\end{equation}
where $P$ is the noise probability, as described previously. The
right-hand side of Eq.~\ref{eq:dyn_fk} is symmetric in respect to values
of $b_t$ around $1/2$ (as can be seen in Fig.~\ref{fig:maj-dyn}), such
that the dynamics for values of $b'_t > 1/2$, can be obtained from
$b'_t= 1 - b_t$, with $b_t < 1/2$. Thus, without loss of generality,
we will only consider the case $b_t\le 1/2$ throughout the paper.

Given any initial starting value $b_0 \le 1/2$, the dynamics will always
lead to a fixed point $b^* \le 1/2$, which is a solution of
Eq.~\ref{eq:dyn_fk}, with $b_{t+1}=b_t\equiv b^*$. This is in general a
solution of a polynomial of order $k$, for which there are no general
closed-form expression. However, since the right-hand side of
Eq.~\ref{eq:dyn_fk} is a monotonically increasing function on $b_t$, we
can conclude there can be at most two possible fixed points: $b^* = 1/2$
(ergodic regime) or $b^* < 1/2$ (non-ergodic regime). Furthermore,
considering the right-hand side of Eq.~\ref{eq:dyn_fk} is a convex
function (for $b_t \le 1/2$, as is always assumed), if the fixed point
$b^* = 1/2$ becomes stable, i.e. $\frac{db_{t+1}}{db_t}|_{b^*=1/2} \le
1$, the other fixed point $b^* < 1/2$ must cease to exist, since in this
case $b_{t+1} > b_t$ for any $b_t < 1/2$. Thus, the value of $P$ for
which $b^*=1/2$ becomes a stable fixed point marks the transition from
non-ergodicity to ergodicity. In order to obtain this value, we need to
compute the the derivative of the right-hand side of Eq.~\ref{eq:dyn_fk}
in respect to $b_t$. Using the derivative of Eq.~\ref{eq:maj} (see
~\cite{evans_maximum_2003} for a detailed derivation of this
expression),
\begin{equation}\label{eq:mdev}
  m'_k(b) \equiv \frac{dm_k(b)}{db} = \frac{k}{2^{k-1}} {k-1 \choose \floor{k/2}} [1-(1-2b)^2]^{\floor{k/2}}
\end{equation}
we have that $(1-2P^*)m'_k(1/2) = 1$, where $P^*$ is the critical value
of noise. Thus, a full expression for $P^*$ is given by
\begin{equation}\label{eq:p_fk}
  P^* = \frac{1}{2} - \frac{2^{k-2}}{k {k-1 \choose \floor{k/2}}}.
\end{equation}
Taking the limit $k\gg 1$, one obtains $P^* \approx \frac{1}{2} -
\frac{1}{2}\sqrt{\frac{\pi}{2k}}$ using the Stirling
approximation. Eq.~\ref{eq:p_fk} is the main result
of~\cite{evans_maximum_2003}. We note however that a slightly less
explicit but more general expression was derived previously
in~\cite{huepe_dynamical_2002}, for the case where the majority function
accepts inputs with different weights.

For a given value of $k$, the value of $b^*$ increases continuously with
$P$ until it reaches $1/2$ for $P\ge P^*$ (see Fig.~\ref{fig:maj-dyn}),
characterizing a second-order phase transition. One can go further and
obtain the critical exponent of the transition by expanding
Eq.~\ref{eq:maj} near $b=1/2$,
\begin{equation}
  m_k(b) = \frac{1}{2} - \frac{1}{2}m'_k(1/2)(1-2b) + \frac{1}{6}\floor{k/2}m'_k(1/2)(1-2b)^3 + O\left((1-2b)^5\right)
\end{equation}
and using it in~\ref{eq:dyn_fk}, and solving for $b^*=b_{t+1}=b_{t}$,
which leads to
\begin{equation}\label{eq:exp_fk}
  b^* \approx \frac{1}{2} - \left[\frac{3}{2}\frac{m'_k(1/2)^3}{\floor{k/2}} \widetilde{P}\right]^{1/2}
\end{equation}
where $\widetilde{P} = P^* - P$. From this expression it can be seen
that the critical exponent is $1/2$, corresponding to the mean-field
universality class.

% \begin{figure}[hbt!]
%   \includegraphics*[width=0.49\columnwidth]{maj-map.eps}
%   \includegraphics*[width=0.49\columnwidth]{maj-dev-map.eps}
%   \caption{Probability $m_k(b)$ of the output of a majority function
%     with $k$ inputs being $1$, if the inputs are $1$ with probability
%     $b$, as given by Eq.~\ref{eq:maj} (left), and its derivative
%     $m'_k(b)$, given by Eq.~\ref{eq:mdev} (right).\label{fig:maj-map}}
% \end{figure}

The values of $b^*$ and $P^*$ can be understood as general bounds on the
minimum error level and maximum tolerable noise, respectively, which
must hold for random networks composed of functions with the same number
of inputs. These are rather stringent conditions, and it is possible to
imagine interesting situations where they are not fulfilled. Therefore,
for more general bounds, one needs to relax these restrictions. We
proceed in this direction in the following section, where we consider
the case of arbitrary in-degree distributions, but otherwise random
connections among the nodes.

\subsection{Arbitrary in-degree distributions}\label{sec:deg-dist}

We turn now to uncorrelated random networks with an arbitrary
distribution of inputs per node (in-degree), $p_k$. Here it is assumed
that the inputs of each function are randomly chosen among all
possibilities, and that the in-degree distribution $p_k$ provides a
complete description of the network ensemble. This configuration was
also considered in~\cite{aldana_phase_2004}, for a more general case
where the inputs can have arbitrary weights. We analyse here the special
case with no weights in more detail, and obtain more explicit results.

The annealed approximation can be used in the same manner as in the
previous section: One considers simply that at each time step the inputs
of each function are randomly chosen~\footnote{Note that this input
  ``rewiring'' has no effect on the in-degree distribution.}. The time
evolution of $b_t$ now becomes,
\begin{equation}\label{eq:dyn_pk}
  b_{t+1} = \sum_k p_k m_k\left((1-2P)b_t + P\right).
\end{equation}
Like for Eq.~\ref{eq:dyn_fk}, there are only two fixed points $b^*\le
1/2$, and the transition can be obtained by analysing the stability of
the fixed point $b^*=1/2$. In an entirely analogous fashion to
Eq.~\ref{eq:p_fk}, using the derivative of the right-hand side of
Eq.~\ref{eq:dyn_pk} one obtains the following expression for the
critical value of noise,
\begin{equation}\label{eq:p_pk}
  P^* = \frac{1}{2} - \left[\sum_kp_k\frac{k {k-1 \choose \floor{k/2}}}{2^{k-2}}\right]^{-1}.
\end{equation}
Considering the limit where all $k\gg 1$, one has $P^* \approx
\frac{1}{2} -
\left[\sqrt{\frac{8}{\pi}}\sum_kp_k\sqrt{k}\right]^{-1}$. Note that the
above expression only holds if $p_k = 0$ for every $k$ which is even, as
is assumed throughout the paper. The critical exponent can also be
calculated in an analogous fashion, and is always $1/2$, unless $p_k$
has diverging moments. In this case the critical exponents will depend
on the details of the distribution (see~\cite{aldana_phase_2004} for a
more thorough analysis).

\begin{figure}[hbt!]
  \begin{center}
  \includegraphics*[width=0.49\columnwidth]{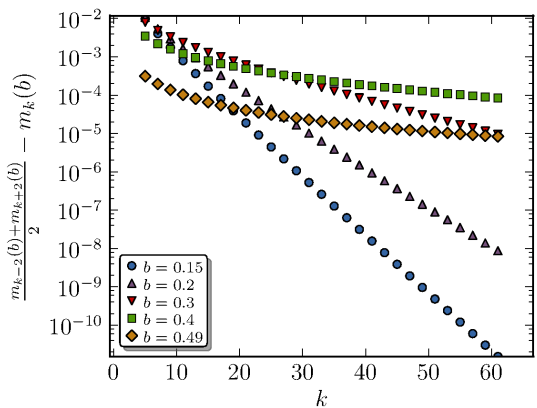}
  \end{center}
  \caption{Convexity of $m_k(b)$, as stated in
    Eq.~\ref{eq:convexity}.\label{fig:maj-convexity}}
\end{figure}

With this result in mind, one can ask the following question: What is
the best in-degree distribution, for a given average in-degree
$\avg{k}$, such that either $b_t$ is minimized or $P^*$ is maximized? As
it will now be shown, in either case the best distribution is the
single-valued distribution, already considered in the previous
section. For simplicity, let us consider the case where $\avg{k}$ is
discrete and odd. We begin with the analysis of $b_t$. We can observe
that for $b\le 1/2$, $m_k(b)$ is a convex function on $k$ (see
Fig~\ref{fig:maj-convexity}),
\begin{equation}\label{eq:convexity}
  m_k(b) \le \frac{m_{k-2}(b) + m_{k+2}(b)}{2},
\end{equation}
and thus by Jensen's inequality we have that $m_{\avg{k}}(b) \le
\avg{m_k(b)}$. Since the equality only holds only for the single-valued
distribution $p_k=\delta_{k,\avg{k}}$ (assuming $b\notin\{0,1/2\}$), the
right-hand side of Eq.~\ref{eq:dyn_pk} will always be larger for any
other distribution $p_k$. The same argument can be made for the value of
$P^*$: Since we have that $(1-2P^*)^{-1} = \sum_kp_km'_k(1/2)$, and
$m'_k(1/2)$ is a concave function on $k$,
\begin{align}
  \frac{m'_{k-2}(1/2) + m'_{k+2}(1/2)}{2} &=
        \frac{m'_k(1/2)}{2}\left[\frac{k-1}{k} + \frac{k+2}{k+1}\right] \\
          &= m'_k(1/2) \left[1 - \frac{1}{2k(k+ 1)}\right] \\
          &<  m'_k(1/2)
\end{align}
we have that $m'_{\avg{k}}(1/2) \ge \avg{m'_k(1/2)}$. Again, the
equality only holds only for $p_k=\delta_{k,\avg{k}}$, which is
therefore the optimal scenario.%
\footnote{Of course, this argument does not hold if $\avg{k}$ is not
  discrete and odd, since in this case the distribution cannot be
  single-valued. But the above argument should make it sufficiently
  clear that in this case the optimal distribution should also be very
  narrow, and similar to the single-valued distribution.}

\begin{figure}[hbt!]
  \begin{center}
    \includegraphics*[width=0.6\columnwidth]{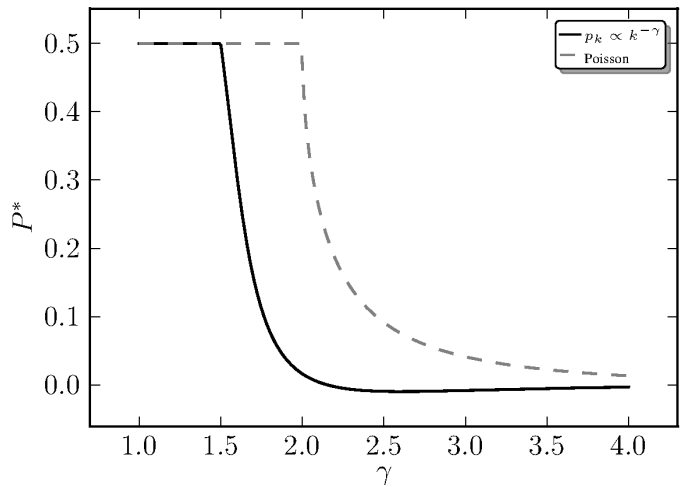}
  \end{center}
  \caption{Critical value of noise $P^*$ as a function of $\gamma$ for
    the scale-free in-degree distribution given by Eq.~\ref{eq:zipf} and
    for the Poisson distribution given by Eq.~\ref{eq:poisson}, where
    $\lambda$ is chosen such that the average in-degree is the same for
    both distributions.\label{fig:pc-zipf}}
\end{figure}

One special case which merits attention is the scale-free in-degree
distribution
\begin{equation}\label{eq:zipf}
  p_k \propto k^{-\gamma},
\end{equation}
which occurs often in many systems, including, as some suggest, gene
regulation~\cite{maslov_computational_2005}. It is often postulated that
networks with such a degree distribution are associated with different
types of robustness, due to their lower percolation
threshold~\cite{cohen_resilience_2000} which can be interpreted as a
resilience to node removal ``attacks''. However, in the case of
robustness against noise Eq.~\ref{eq:zipf} by itself does not confer any
advantage. For instance, from Eq.~\ref{eq:p_pk}, using Stirling's
approximation one sees that the expression within brackets will diverge
only if $\gamma \le 3 / 2$, leading to $P^* = 1/2$.  This means that for
$3/2 < \gamma \le 2$, we have that the average in-degree diverges
($\avg{k}\to\infty$) but the critical value of noise is still below
$1/2$. This is considerably worse, for instance, than a fully random
network with in-degree distribution given by a slightly modified
Poisson, which is defined only over odd values of $k$,
\begin{equation}\label{eq:poisson}
  p_k = \frac{1}{\sinh\lambda}\frac{\lambda^k}{k!},
\end{equation}
with $\avg{k} = \lambda/\tanh\lambda$. For this distribution, we have
that $P^* \to 1/2$ for $\avg{k}\to\infty$, as one would expect also for
the single-valued distribution. A comparison between these two
distributions is shown in Fig.~\ref{fig:pc-zipf}.

The above analysis shows that the single-valued in-degree distribution
$p_k=\delta_{k,\avg{k}}$ is the best one can hope for with a given
average in-degree $\avg{k}$, as long as the inputs of each function are
randomly chosen. However, this is a restriction which does not need be
fulfilled in general.  In order to obtain more general bounds, one needs
to depart from this restriction, and consider more heterogeneous
possibilities, which is the topic of the next section.

\subsection{Arbitrary topology:  Stochastic blockmodels}\label{sec:block}

We now consider a much more general class of networks known as
stochastic
blockmodels~\cite{holland_stochastic_1983,faust_blockmodels:_1992,boguna_class_2003},
where it is assumed that every node in the network can belong one of $n$
distinct classes or ``blocks''. Every node belonging to the same block
has on average the same characteristics, such that we need only to
describe the degrees of freedom associated with the individual
blocks. In particular we use the degree-corrected
variant~\cite{karrer_stochastic_2011} of the traditional stochastic
blockmodel, which incorporates degree variability inside the same block.
Here, we define $w_i$ to be the fraction of the nodes in the network
which belong to block $i$, and $p^i_k$ is the in-degree distribution of
block $i$. The matrix $w_{j\to i}$ describes the fraction of the inputs
of block $i$ which belong to block $j$. We have therefore that
$\sum_iw_i=1$, $\sum_jw_{j\to i}=1$ and $\sum_{i,k}kw_ip^i_k =
\avg{k}$. Since the out-degrees are not explicitly required to describe
the dynamics, they will be assumed to be randomly distributed, subject
only to the restrictions imposed by $w_i$ and $w_{j\to i}$.

In the limit where the number of vertices $Nw_i$ belonging to each
blocks $i$ is arbitrary large, we can use a modified version of the
annealed approximation to describe the dynamics: Instead of randomly
re-assigning inputs for each function, we choose randomly only amongst
those which do not invalidate the desired block structure. In other
words, we impose that after each random input rewiring, the inter-block
connections probabilities are always given by $w_{j\to i}$. In this way,
we maintain the dynamic correlations associated with the block
structure, and remove those arising from quenched topological
correlations present in a single realization of the blockmodel
ensemble. Due to the self-averaging properties of this ensemble, for
sufficiently large networks the annealed approximation is expected to be
exact, in the same way it is for random networks without block
structures.

With this ansatz, we can write the average value of $b_i$ for each block
over time as
\begin{equation}\label{eq:dynblock}
  b_i(t+1) = \sum_kp_k^im_k\left((1-2P)\sum_j w_{j\to i}b_j(t) + P\right),
\end{equation}
which is a system of $n$ coupled maps. It is easy to see that
$b^*_i=1/2$ is a fixed point of Eq.~\ref{eq:dynblock}. In order to
perform the stability analysis we have to consider the Jacobian matrix
of the right-hand side of Eq.~\ref{eq:dynblock},
\begin{equation}
  J_{ij} = \frac{\partial b_i(t+1)}{\partial b_j(t)} = (1-2P) w_{j\to i} \sum_kp^i_km'_k\left((1-2P)\sum_j w_{j\to i} b_j(t)\right).
\end{equation}
At the fixed-point $b_i(t) = 1/2$ we can write the Jacobian as
\begin{equation}
  \bm{J^*} = (1-2P)\bm{M},
\end{equation}
where matrix $\bm{M}$ is given by
\begin{equation}\label{eq:M}
  M_{ij} = w_{j\to i} \sum_kp^i_km'_k(1/2).
\end{equation}
The largest eigenvalues of $\bm{J^*}$ and $\bm{M}$, $\lambda$ and $\xi$
respectively, are related to each other simply by $\lambda =
(1-2P)\xi$. Since the fixed-point in question will cease to be stable
for $\lambda = 1$, we have that the critical value of noise is given by
\begin{equation}\label{eq:pcblock}
  P^* = \frac{1}{2} - \frac{1}{2\xi}.
\end{equation}
Thus, for $P > P^*$ the fixed point $b_i(t) = 1/2$ becomes a stable
fixed-point, and this marks the transition from non-ergodicity to
ergodicity, as in the previous cases.

We note that the sizes of the blocks $w_i$ play no role in
Eq.~\ref{eq:pcblock}, and only the correlation probabilities $w_{i\to
  j}$ and the in-degree distributions $p^i_k$ define the value of
$P^*$. For this reason, the average error $b^* = \sum_i w_i b_i$ on the
network may not be always a suitable order parameter to identify the
aforementioned phase transition, since the blocks which are responsible
for the value of $P^*$ may be arbitrarily small in comparison to the
rest of the network. However, these are obviously corner cases, since
the most interesting situations are those where all blocks are relevant
to the dynamics (or a given block could be otherwise ignored).

Given any desired many-block structure, one could find the largest
eigenvalue $\xi$ of the matrix $\bm{M}$ and then determine the critical
value of noise with Eq.~\ref{eq:pcblock}. In the following, we will
focus on the simplest nontrivial block structure which is composed only
of two blocks. Such 2-block systems are fully accessible analytically,
and are sufficient to obtain more general upper and lower bounds on the
values of $P^*$ and $b^*$, respectively.

\subsection{2-block structures}

Here we consider networks composed of two blocks, where the block with
the largest average in-degree will be labeled ``core''. The size and
average in-degree of the core block are $w_c$ and $k_c$ respectively,
and for the non-core block $w_r = 1-w_c$ and $k_r=(\avg{k} -
w_ck_c)/(1-w_c)$. For simplicity, we will consider that the in-degree
distribution of each block is the single-valued distribution $p^i_k =
\delta_{k,k_i}$, where $k_i$ is the average in-degree of the block.

\begin{figure}[hbt!]
  \begin{center}
    \begin{minipage}[b]{0.3\columnwidth}
      \centering
      \includegraphics*[scale=0.45]{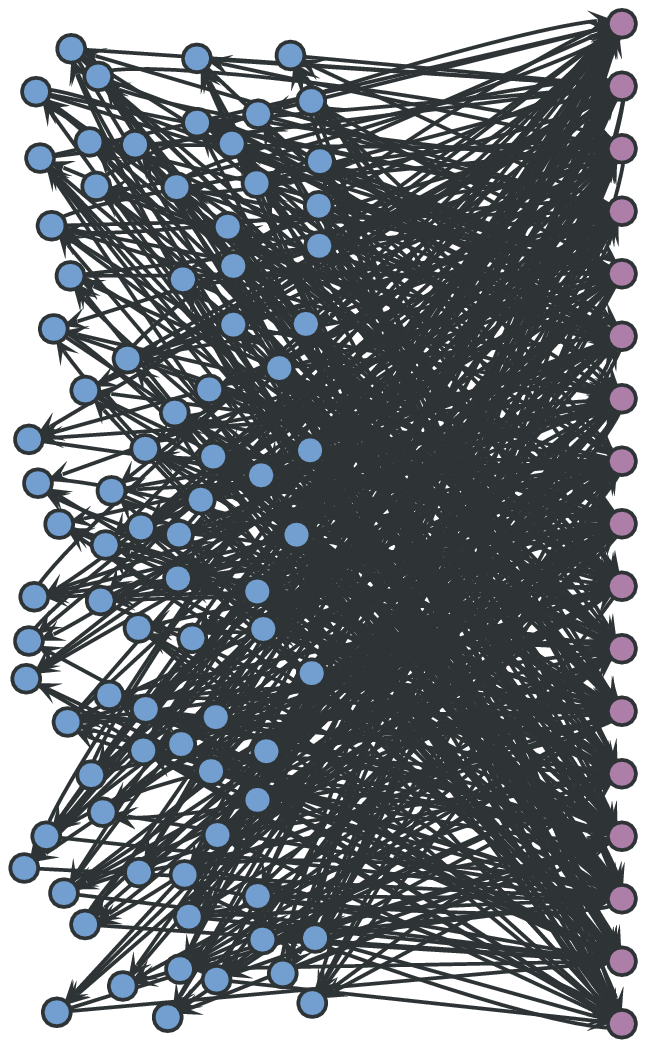}\\
      $a=0$\\
      Restoration
    \end{minipage}
    \begin{minipage}[b]{0.3\columnwidth}
      \centering
      \includegraphics*[scale=0.45]{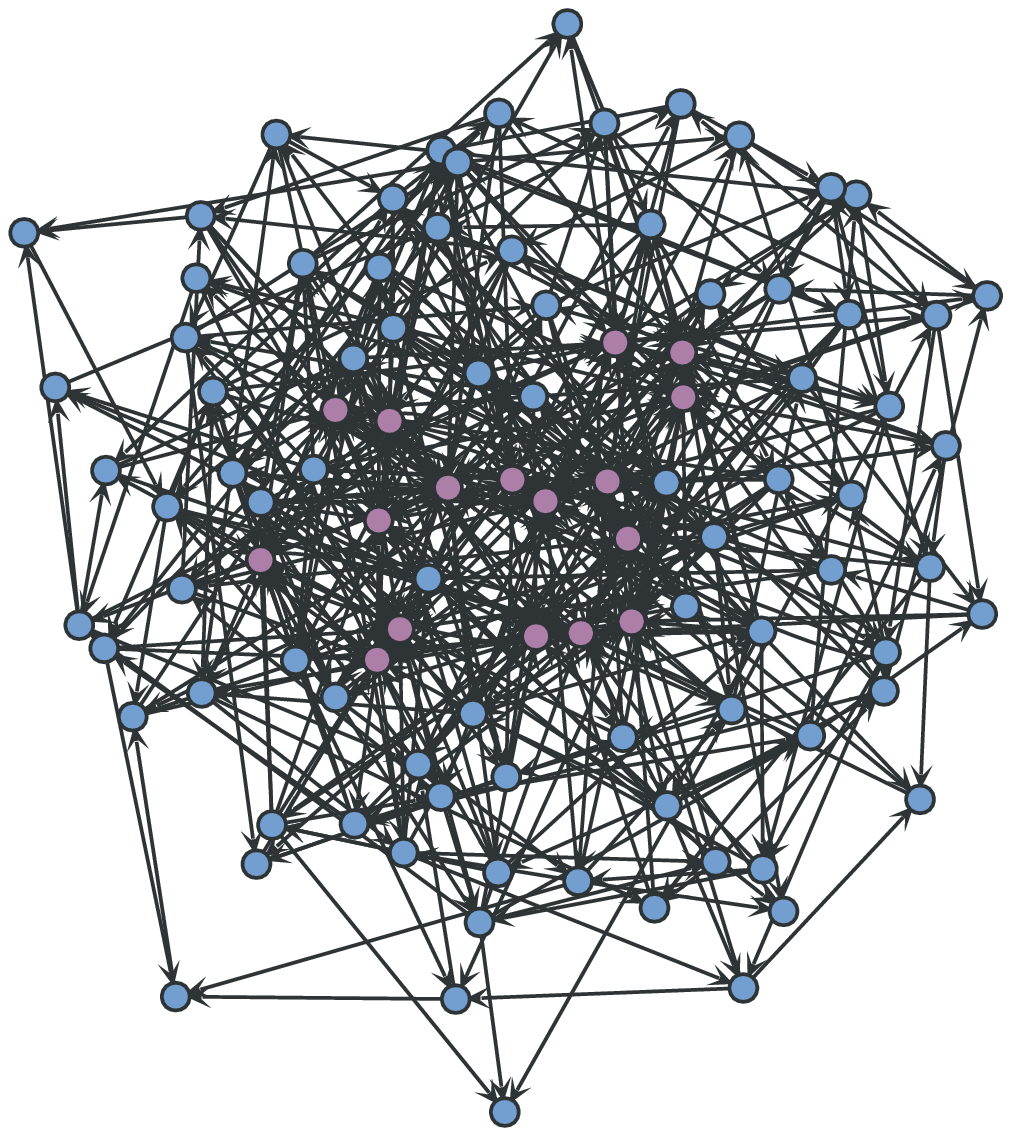}\\
      $a=1/2$\\
      Random
    \end{minipage}
    \begin{minipage}[b]{0.3\columnwidth}
      \centering
      \includegraphics*[scale=0.45]{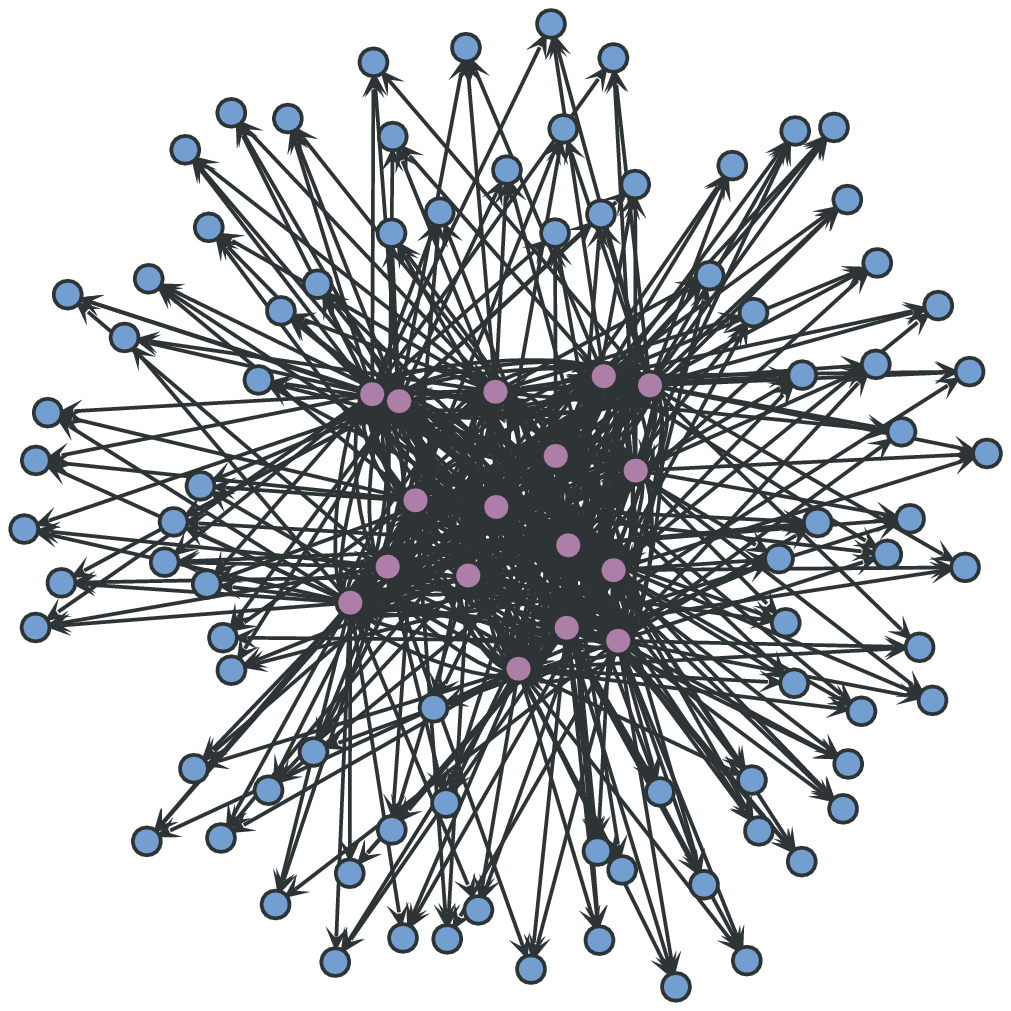}\\
      $a=1$\\
      Segregated core
    \end{minipage}
    \end{center}
    \caption{Three distinct 2-block structures possible with the
      parametrization given by Eq.~\ref{eq:aparm}, for different values
      of the parameter $a$.\label{fig:block}}
\end{figure}

The matrix $w_{j\to i}$ has the general form
\begin{equation}
  \bm{w_{\to}} =
 \left(
    \begin{array}{cc}
      w_{c\to c} & w_{c\to r} \\
      w_{r\to c} & w_{r\to r}
    \end{array}
  \right)
  =
  \left(
    \begin{array}{cc}
      m_c & m_r \\
      1-m_c & 1 - m_r
    \end{array}
  \right),
\end{equation}
with only two free variables $m_c$ and $m_r$, denoting the fraction of
inputs which belong to the core block, for both blocks. Instead of
considering all possible values of $m_c$ and $m_r$, we consider the
following parametrization
\begin{equation}\label{eq:aparm}
\begin{aligned}
    m_c &=
    \begin{cases}
      4 a (1 - a) w_c & \text{if } a \le 1/2 \\
      m_r             & \text{if } a > 1/2
    \end{cases}\\
    m_r &= 1 - 4a(1 - a)(1 - w_c),
  \end{aligned}
\end{equation}
where the single parameter $a \in [0,1]$ allows for the topology to be
continuously varied between three distinct topological configurations
(see Fig.~\ref{fig:block}): For $a=0$ we have a ``restoration''
topology, where the network is bipartite, and all inputs from the
non-core block belong to the core block and vice-versa; for $a=1/2$ the
inputs are randomly selected; and for $a=1$ we have a ``segregated
core'' structure, where all the inputs of both blocks belong exclusively
to the core block.

For this system we can write the matrix $\bm{M}$ from Eq.~\ref{eq:M} as
\begin{equation}
  \bm{M} =
 \left(
    \begin{array}{cc}
      w_{c\to c} m'_{k_c}(1/2) & w_{r\to c} m'_{k_c}(1/2) \\
      w_{c\to r} m'_{k_r}(1/2) & w_{r\to r} m'_{k_r}(1/2)
    \end{array}
  \right),
\end{equation}
from which we can extract the largest eigenvalue $\xi$,
\begin{multline}\label{eq:2xi}
  \xi = \frac{1}{2} (w_{c\to c} m'_{k_c}(1/2) + w_{r\to r} m'_{k_r}(1/2))
  \quad + \\
  \frac{1}{2}\sqrt{4 w_{r\to c} w_{c\to r}m'_{k_c}(1/2)m'_{k_r}(1/2) + 
    \left(w_{c\to c} m'_{k_c}(1/2) - w_{r\to r} m'_{k_r}(1/2)\right)^2}.
\end{multline}
From $\xi$, the critical value of noise can be obtained by
Eq.~\ref{eq:pcblock}.

\begin{figure}[hbt!]
  \includegraphics*[width=0.49\columnwidth]{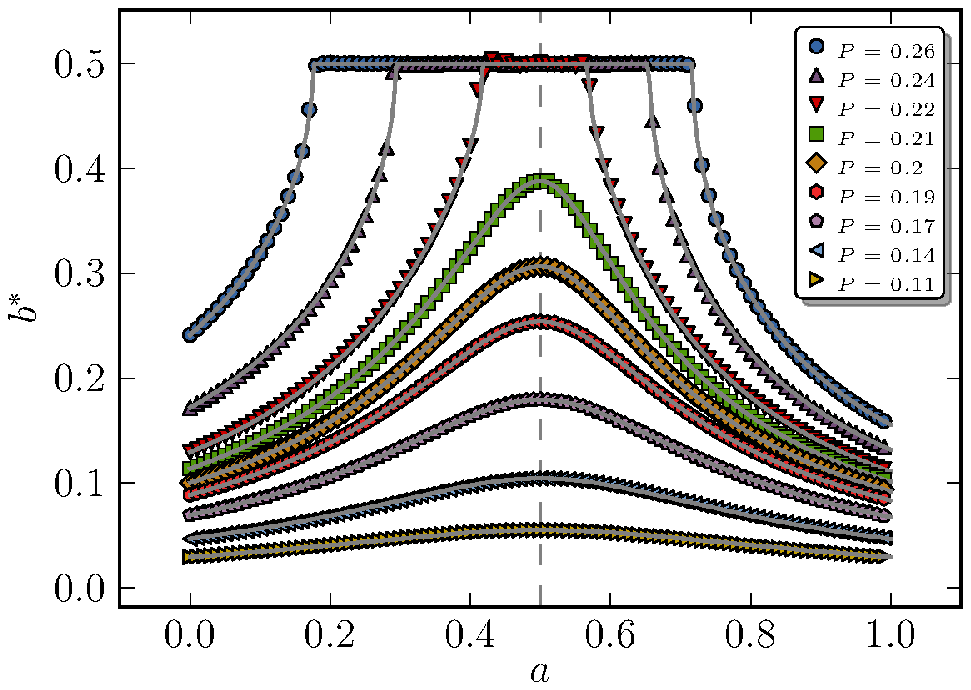}
  \includegraphics*[width=0.49\columnwidth]{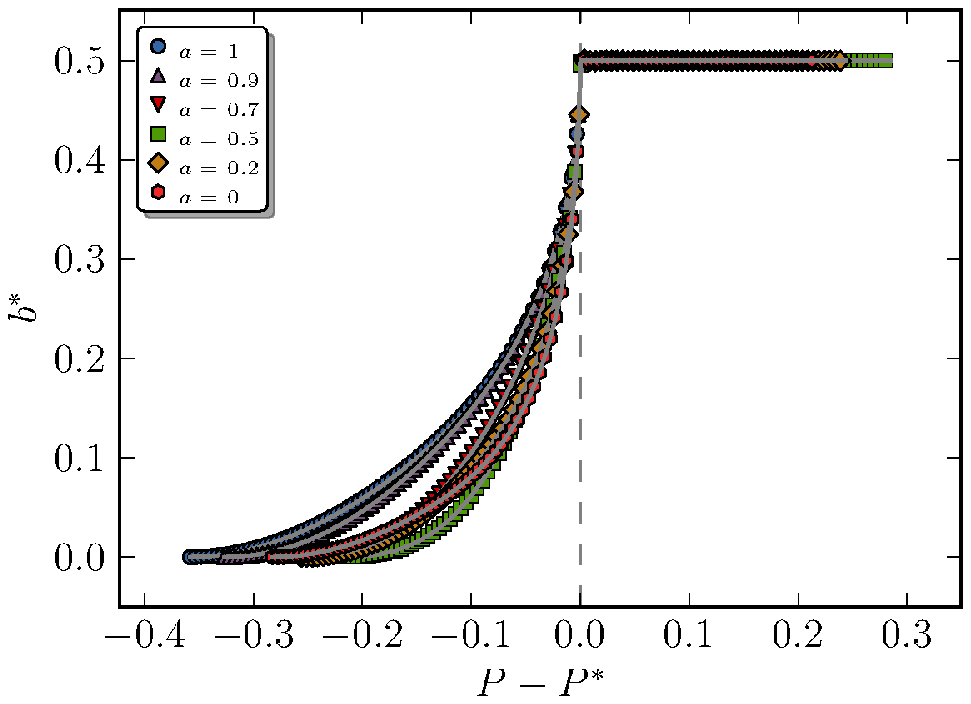}
  \caption{Average error $b^*$ as a function of $a$ for different values
    of noise $P$ (left) and as a function of the reduced noise $P-P^*$,
    with $P^*$ computed according to Eqs.~\ref{eq:2xi}
    and~\ref{eq:pcblock}, for several values of $a$ (right). All curves
    are for $\avg{k}=5$, $k_r=3$ and $k_c=19$. The symbols are results
    of numerical simulations of quenched networks with $N=10^5$ nodes,
    and the solid lines are numerical solutions of
    Eq.~\ref{eq:dynblock}.\label{fig:b}}
\end{figure}

\begin{figure}[hbt!]
  \includegraphics*[width=0.49\columnwidth]{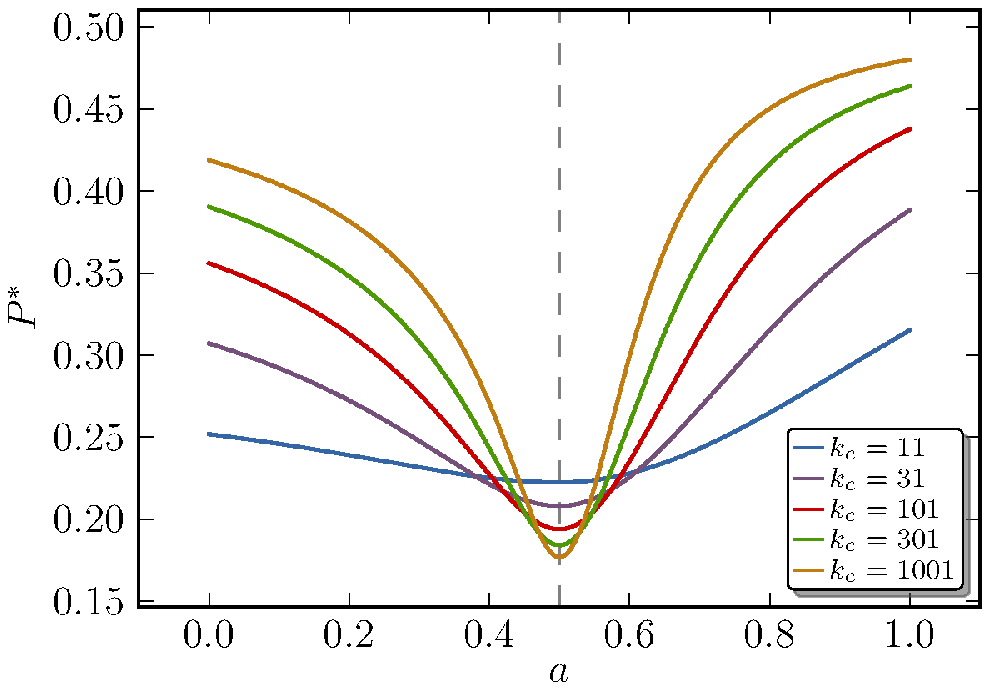}
  \includegraphics*[width=0.49\columnwidth]{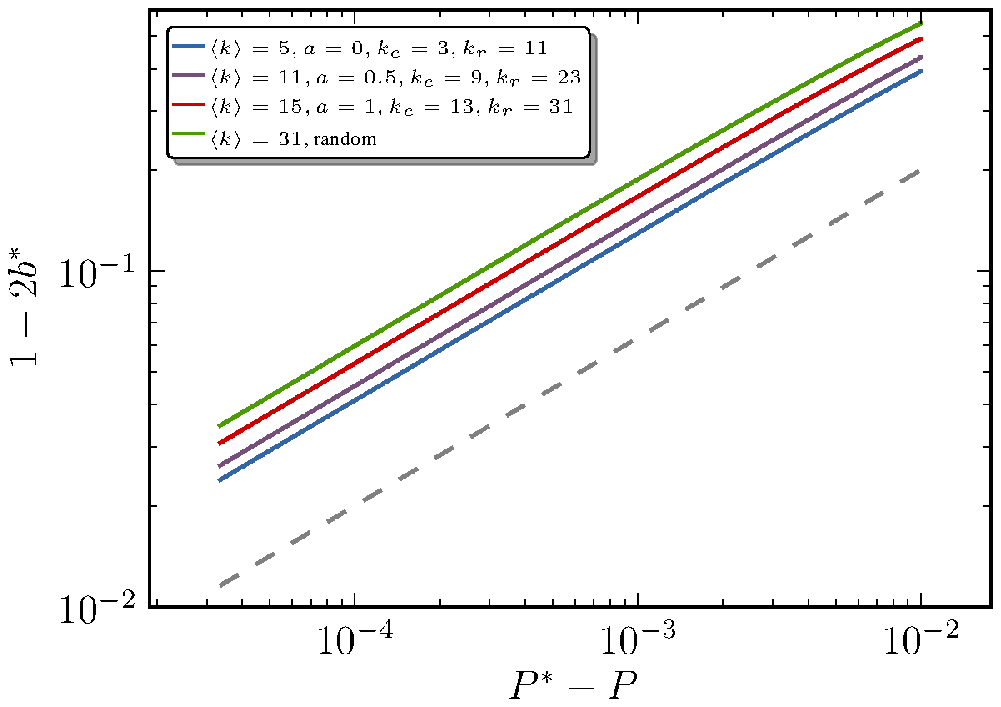}
  \caption{Critical value of noise $P^*$ as a function of $a$, for
    several values of $k_c$, with $k_r=3$ and $\avg{k}=5$ (left), and
    value of $1-2b^*$ as a function of $P-P^*$ close to the critical
    point, for different values of $\avg{k}$, $a$, $k_r$ and $k_c$
    (right). The dashed line corresponds to a function proportional to
    $(P-P^*)^{1/2}$.\label{fig:pc}}
\end{figure}

The general behaviour of the asymptotic average error $b^* \equiv
\lim_{t\to\infty}\avg{b_i(t)}$, computed from Eq.~\ref{eq:dynblock} as a
function of $a$ is shown in Fig.~\ref{fig:b} for $\avg{k} = 5$ and $k_r
= 3$, and several values of $k_c$ (and $w_c$ chosen accordingly). In the
same figure are shown results from numerical simulations of quenched
networks with $N=10^5$ nodes, evolved according to Eq.~\ref{eq:bn_dyn},
showing perfect agreement. On the right of Fig.~\ref{fig:b} are shown
the values of $b^*$ according to the reduced noise $P-P^*$, with $P^*$
computed according to Eqs.~\ref{eq:2xi} and~\ref{eq:pcblock}. The
calculated values of $P^*$ for several values of $k_c$ are plotted on
the right of Fig.~\ref{fig:pc}. The nature of the phase transition is
systematically the same, as can be seen in the right of
Fig.~\ref{fig:pc}, where the slope of the curves correspond to
mean-field critical exponent $1/2$.

It is interesting to compare the performance of the restoration ($a=0$)
and segregated core ($a=1$) topologies. Both outperform the random
topology ($a=1/2$), but the segregated core is always the best possible,
having both the lowest values of $b^*$ and largest values of $P^*$. This
is not surprising, since the segregated core is nothing more than an
isolated network, which is more densely connected than the whole
network, to which the remaining nodes are enslaved. On the other hand it
is rather interesting how the restoration topology ($a=1$) is only
marginally worse than the segregated core, since in this situation every
node is dynamically relevant.  We note that the relative advantage of
the partially random topologies ($0<a<1$) may depend on the actual value
of noise. This can be seen in Fig.~\ref{fig:b} (right), where the curves
for $b^*$ with different values of $a$ cross each other when $P-P^*$ is
varied (the same is also observed when the curves are plotted against
$P$). The reason for this is that the relative advantage of the
segregated core topology in respect to restoration may manifest itself
only as the value of noise approaches the critical point. For lower
values of noise it is possible, for instance, for a full restoration
topology with $a=0$ to outperform a partial segregated core structure
with $a=0.9$, since it will perform comparably to a full segregation,
$a=1$ (see Fig.~\ref{fig:b}, left). However, as noise is increased the
relative advantage of the segregated topology makes up for this
difference. In the general case, therefore, the optimal topology will
depend on the value of noise.

\begin{figure}[hbt!]
  \includegraphics*[width=0.49\columnwidth]{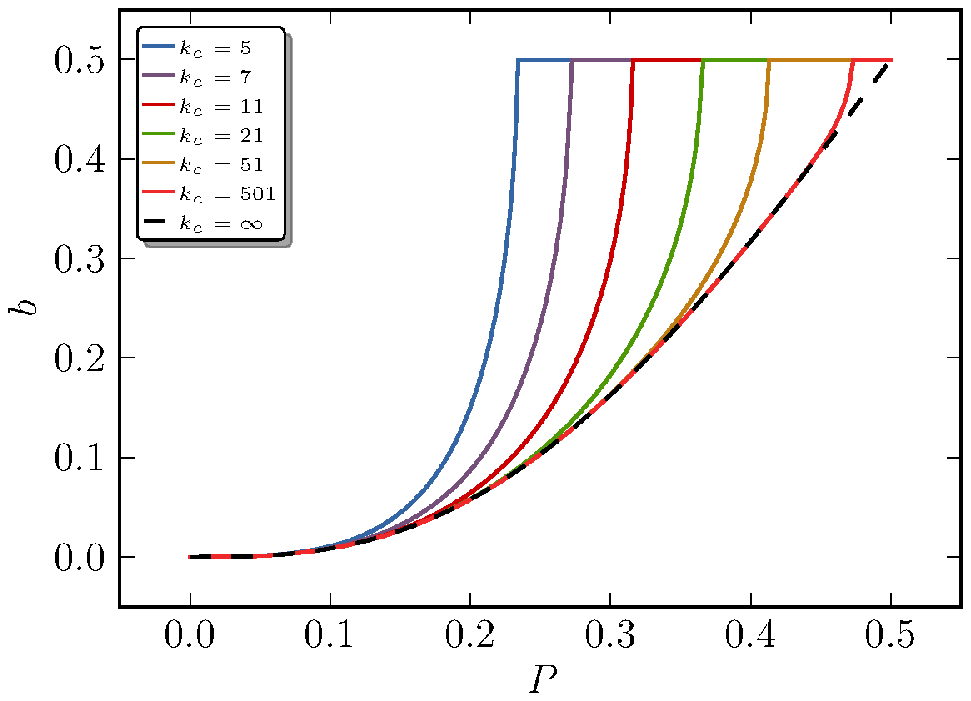}
  \includegraphics*[width=0.49\columnwidth]{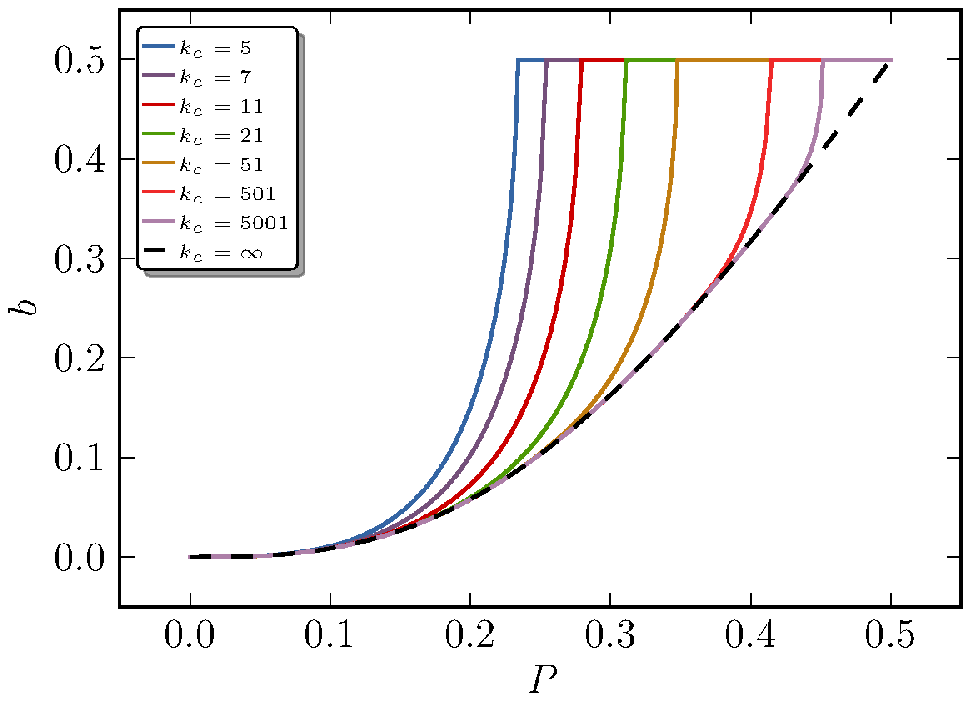}
  \caption{Values of $b^*$ as a function of $P$ for 2-block structures
    with $a=1$ (left) and $a=0$ (right), with $w_c= 1 / (100 \times k_c)$,
    $k_r=5$ and several values of $k_c$. The dashed curves are given by
    Eq.~\ref{eq:bo} with $\avg{k} = k_r$.\label{fig:optimum}}
\end{figure}

Either with the restoration and segregated core topologies, the values
of $b^*$ and $P^*$ become increasingly better for larger values of
$k_c$, as can be seen in Figs.~\ref{fig:pc} and~\ref{fig:optimum}. One
can therefore postulate that an optimum bound can be achieved for
$k_c\to\infty$. Let us consider the situation where $w_c \propto 1 /
k_c$, such that $\lim_{k_c\to\infty}\avg{k} = k_r$. For both $a=0$ and
$a=1$ the value of $b^*$ approaches asymptotically $m_{\avg{k}}(P)$, for
$k_c\to\infty$, as can be seen in Fig.~\ref{fig:optimum}. This means
that the average error of the core nodes will eventually vanish, and the
remaining nodes will encounter the optimal scenario where the inputs are
affected by the noise $P$ alone, and the error does not accumulate over
time. It is therefore safe to conclude that
\begin{equation}\label{eq:bo}
  b_{\text{min}} = m_{\avg{k}}(P)
\end{equation}
is a general lower bound on the average error on a network with average
in-degree $\avg{k}$ and an arbitrary topology, which is asymptotically
achieved for both the restoration and segregation topologies, for
$k_c\to \infty$.

\section{Conclusion}\label{sec:conclusion}

We have investigated the behaviour of optimal Boolean networks with
majority functions and different topologies in the presence of
stochastic fluctuations. The dynamics of these networks undergo a phase
transition from ergodicity to non-ergodicity. The non-ergodic regime can
be can be interpreted as robustness against noise, since there is a
permanent global memory of the initial condition. The ergodic phase, on
the other hand, represents a situation where the effect of noise has
destroyed any possible long-term dynamical organization of the
system. We obtained, both analytically and numerically, the average
error and the critical value of noise for networks composed of arbitrary
in-degree distributions and for a more general stochastic blockmodel,
which can accommodate a wide variety of network structures. We showed
that both the average error level as well as the critical value of noise
are improved both for the segregated core and restoration topologies,
where the dynamics is dominated by a smaller subset of nodes, which have
an above-average in-degree. In the limit where the average in-degree of
these ``core'' nodes diverges, the network achieves an optimum bound,
which corresponds to the maximum resilience attainable.

In a separate work~\cite{peixoto_emergence_2011}, we show that
segregated core structures emerge naturally out of an evolutionary
process which favors robustness against noise.

As was discussed, the networks considered are made from optimal
elements, which in isolation have the best possible behaviour. Because
of this, the results obtained have a general character, and show the
best scenario which can in general be achieved, under the constraints
considered. However, it is important to point out that there are
different types of stochastic fluctuations which can be considered in
Boolean systems. Other than the type of noise considered in this work,
it is possible for instance to incorporate fluctuations in the
\emph{update schedule} of the nodes~\cite{klemm_topology_2005}. It has
been shown in~\cite{greil_dynamics_2005}, for random networks, that even
if the update schedule is completely random, ergodicity is preserved,
and the dynamics eventually leads to distinct attractors. Furthermore,
it was shown in~\cite{peixoto_boolean_2009} that it is possible to
obtain absolute resilience against noise in the update sequence, where
the trajectories are always the same, independent of the update schedule
used. In~\cite{schmal_boolean_2010} this type of resilience has been
coupled with single-flip perturbations, which correspond to very small
values of the noise parameter $P$ considered in this work, and it was
shown that arbitrary mutual resilience is also possible. The broader
question of how a single system can be simultaneously robust against
many different types of perturbations, and which features become more
important in this case, still needs to be systematically tackled.

\section*{References}

%\bibliographystyle{ieeetr}
%\bibliography{bib}

\end{document}